# Searching and Sorting Algorithms for Quantum Annealing Computers

Robert A. Dunn
Lockheed Martin Corporation, USA

Algorithms for searching and sorting data sets on quantum annealing systems are presented. Search algorithms for unordered data sets are developed. A sorting algorithm for data sets is provided, with a consideration of sort stability. Scalability of the algorithms, considering both the number of qubits required and the qubit connectivity, is characterized as a function of problem size.

**Introduction**

Algorithms for searching and sorting data sets are among the first taught to students of computer science due to the widespread applicability of these types of algorithms in problem solving. Grover [1] provided a probabilistic algorithm for searching unordered data sets using gate array quantum computers, but there is a notable gap in the literature concerning searching and sorting algorithms for quantum annealing computers. This paper addresses that gap by providing an initial presentation of algorithms suitable for implementation on quantum annealing and adiabatic quantum computers.

This analysis is not concerned with identifying possible cases in which a quantum algorithm might outperform existing digital search and sort algorithms [2] [3]. Instead, we explore the techniques useful for including search and sort capabilities as components in larger algorithms for application development on quantum annealing systems.

Presented herein are:

- a search algorithm for unordered data sets
- a search algorithm for bounding indices in ordered data sets
- a sort algorithm for unordered data sets

This paper uses the following nomenclature:

Let {A} represent the input data set of N elements, the $i^{th}$ element of which is A[i].

Let {B} represent the output, sorted data set of N elements, the $i^{th}$ element of which is B[i].

Let $K_i$ be the size of the array index variable type (eg 4 bits for arrays up to 16 elements)

Let $K_v$ be the size of the array data element variable type (eg 32 bits for integer data)

Let x be the value sought in a search, of the same type as the array entries

Let n be the index into the data set {A}, such that A[n] = x

We consider cases of fixed-size arrays of N elements, with data elements of size $K_v$ logical qubits.

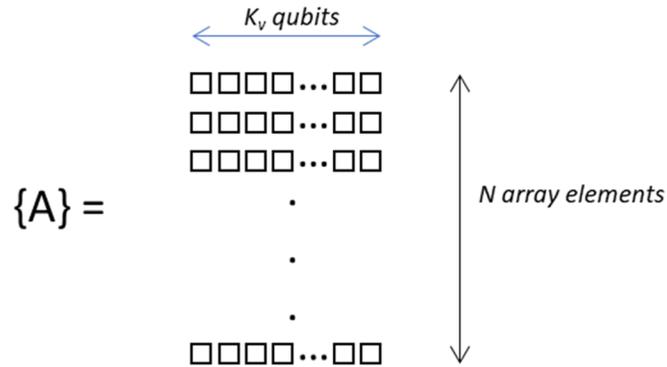

*Figure 1 Logical structure of data array of N elements, each of which is $K_v$ qubits in size.*

The assignment of logical qubits (i.e., variables in the Hamiltonian) to physical qubits (i.e., specific hardware quantum bits in the quantum annealer) takes place during the embedding process, which is dependent on the hardware graph of the quantum annealing computer used. The algorithms presented here make no assumptions concerning the hardware graph connectivity or the proximity of array element qubits in the final embedding.

Of particular interest are cases in which either the data element values are established (in whole or in part) during the run, or in the which the index is established during the run of the quantum program. Those applications for which both the data values and the array indices are known prior to Hamiltonian construction may be efficiently processed on digital computing systems during the creation of the quantum annealing Hamiltonian and thus require no quantum system resources.

The process of constructing the problem Hamiltonian for a given run will be termed 'compilation', for sake of convenience and analogy with traditional programming methodologies.

**Creating Data Arrays in Quantum Annealing Applications**

Because we are considering only the case of fixed-length arrays in the analysis, the problems of dynamic resource allocation are not present. This greatly simplifies the allocation of logical qubit variables to the array variables since the assignment may be performed during compilation.

Assigning a value to a specific array element for a known index may be trivially constructed at compile time because all qubit variables are readily identified. The terms in the assignment portion of the Hamiltonian consist of bitwise equivalence constraints. This is true even when the value is not known at compile time, e.g., constructed as part of the computation.

The more complex case occurs when the index is not known in advance, i.e., is a variable in the quantum algorithm. In this case, the common programming expression for assigning a value to an array element

```
A[n] := x
```

where n is a run-time dependent index, differs in the quantum annealing case from the digital programming case because the superposition of all possible indices is involved. The problem may be rephrased as a constraint by requiring

```
(i == n) and (A[i] == x)
```

evaluated for all elements of the array A. A similar formulation of the problem as

```
if ( i == n ) then
    A[i] := x
end if
```

over all elements in {A} produces equivalent results, albeit with more 3-local terms in the Hamiltonian.

The first variant is also preferable because the symmetry of array assignment and array access implicit in the relationship

```
A[n] == x
```

permits a ready formulation for table lookup using quantum variable indices.

This difference in behavior between the cases of known and unknown index variables for quantum annealing algorithms reveals a key limitation: access and assignment operations using quantum variables as array indices are expensive operations, in that they require relatively highly connected logical qubit graphs. This is in stark contrast to the ease of variable index array access in digital computers accomplished by pointer arithmetic.

**A Search Algorithm for Unordered Data Sets**

The Hamiltonian for the problem containing, in part, the search algorithm is typically composed of several terms corresponding to the data, constraints, etc. Accordingly, the portions of the problem Hamiltonian corresponding to the search implementation naturally decompose into two blocks: the array element comparisons (which are dependent on the array data type and array index type) and the search algorithm proper.

The array element comparisons require $O(K_i + K_v)$ logical qubits, since in the worst case a bitwise comparison is sufficient to confirm equality (it may be less for composite data structures in which only a subset of the array element value fields are used for purposes of comparison). Therefore, this analysis focuses on the resources required for the efficient implementation of the search algorithm rather than the value comparisons.

The two primary scenarios for data search ask some variant of the following questions:
1) Does a given value *x* exist in the data set?
2) Which index *n* in the data set corresponds to the given value *x*?

Mathematically, we seek:

$$n \in [1..N]$$

such that

```
A[ n ] == x
```

if such an n exists, with an additional flag indicating whether such n was found.

We define the 1-qubit variable `not_found` with a value of 0 if such *n* exists, and a value of 1 if no index *n* was found, i.e., *x* was not in the data set. The reason for choosing the negation of the search result as a variable will be evident in the later section on unmatched search scenarios.

For a quantum algorithm, the index *n* and variable *x* are composite quantum variables composed of multiple qubits (e.g., a quantum integer or a string key into a key/value dictionary). For N elements in the input array {A}, the index variable must be at least of size $K_i \geq \lceil log_2 N \rceil$ qubits.

The following qubit variables are defined for each element in the data set {A}:

$$I_i = \begin{cases} 1, & i == n \\ 0, & \text{otherwise} \end{cases}$$

$$V_i = \begin{cases} 1, & A[i] == x \\ 0, & \text{otherwise} \end{cases}$$

Hamiltonian terms for these comparisons are generated by the algorithm shown in Appendix A.

We may now construct a Hamiltonian term

$$H_{search} = \left(1 - \sum_{i=1}^{N} I_i V_i\right)^2$$

*Equation 1 Basic Hamiltonian term for array search.*

which has a global minimum when exactly one element has been selected meeting the search criteria. This also handles arrays in which the target value appears more than once, though the ground states will be degenerate in that case.

For systems supporting at most 2-local terms, 2N ancillary qubits are needed to reduce $H_{search}$ to 2nd order.

To examine the resource expense and scalability of this algorithm, we constructed $H_{search}$ Hamiltonians for arrays of 8-bit quantum integers ($K_v$ = 8) of various sizes and documented the number of qubits required to implement the search algorithm. Figure 2 demonstrates the number of logical qubits required to implement the basic search shown in Equation 1 increases linearly with array size. This includes the search algorithm variables and ancillary qubits, but excludes the qubits actually defining the array {A}.

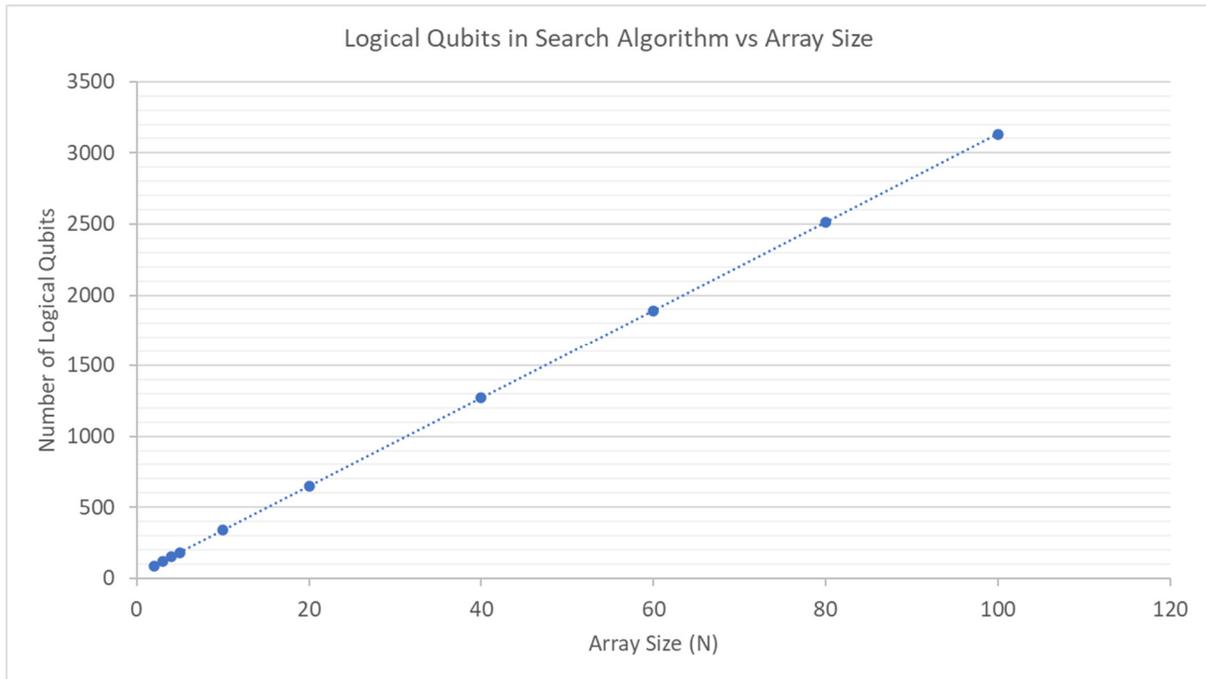

*Figure 2 Logical qubits required to search an array of 8-bit quantum integers, as a function of array size.*

**Preprocessing Effort**

An important consideration when evaluating the feasibility of quantum algorithms is the classical preprocessing effort required. If the preprocessing stage were too computationally expensive, it might render the quantum stage of the application moot. For this search algorithm, the preprocessing consists of creating the Hamiltonian terms to implement the desired functionality, as follows:

The index comparison terms in the Hamiltonian to construct the `I`$_i$ values are, in the worst case, bitwise comparisons taking O($K_i$ N) steps to construct, including order reduction.

Similarly, the terms supporting the population of the `V`$_i$ array element comparison variables require O($K_v$ N) to construct, including order reduction.

The H$_{search}$ term requires O($N^2$) steps for creation and order reduction.

Thus, the preprocessing effort required to construct the Hamiltonian terms to implement this search algorithm scales quadratically with the number of array elements.

**Search Results in Un-matched Scenarios**

To handle the case in which the sought value is not present in the data, we construct a Hamiltonian term such that the energy values meet the following condition:

(energy of successful search) < (energy of unsuccessful search) < (energy of invalid logic)

We require that the energy of unsuccessful searches be larger than that of successful searches to avoid the 'lazy search' case – we accept the answer of 'not found' only when all valid comparisons were tried and failed. This is accomplished in two steps – a modification of the search Hamiltonian term defined in Equation 1, and a small energy penalty associated with the `not_found` variable. To avoid invalidating the comparison logic, the energy state of the failed search must be lower than the energy of the lowest non-ground state of the successful search term. Thus,

$$H_{search} = \left(1 - not\_found - \sum_{i=1}^{N} I_i V_i\right)^2 + \frac{1}{2} not\_found$$

*Equation 2 Search Hamiltonian term, including option for search failure.*

### Reducing the Qubit Connection Graph

The search QUBO formed from Equation 2 above has a qubit graph with a connectivity too high to be conveniently accommodated by current hardware for arrays of more than 100 elements, as seen in Figure 3.

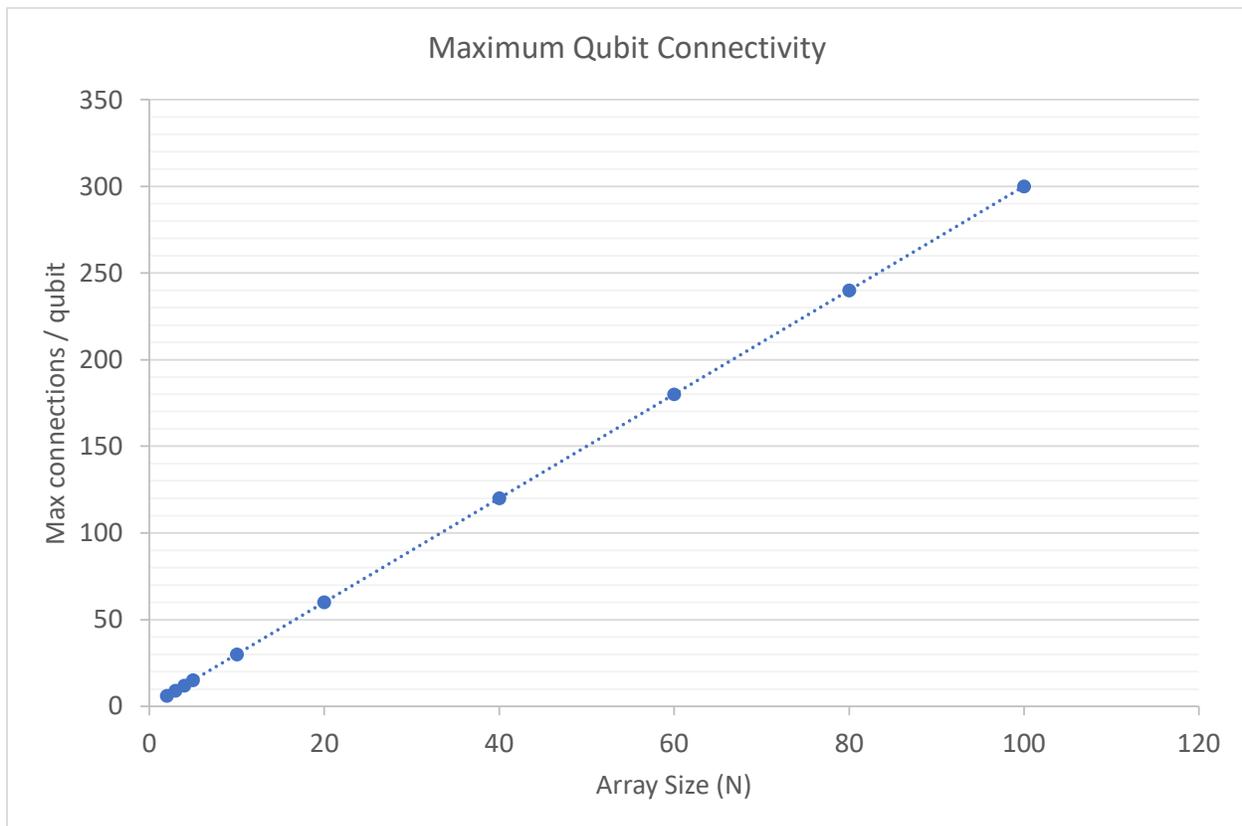

*Figure 3 Maximum number of connections for qubits in search Hamiltonian, as a function of data array size.*

The maximum connectedness of the qubit graph from the search Hamiltonian grows linearly with the number of array elements, specifically there are 3N qubit connections for each bit of the sought variable `x`. The number of logical qubits with high connectedness also grows linearly with the array size, resulting in a difficult hardware embedding.

The overall connectedness of the qubit graph can be reduced by rephrasing the $H_{search}$ term from a summation to a logical expression. The construction of an intermediate variable indicating whether any solution was found may be used as follows:

$$found = \bigvee_{i=1}^{N} I_i V_i$$

and

$$H_{search} = (1 - not\_found - found)^2 + \frac{1}{2} not\_found$$

*Equation 3 Search Hamiltonian term, with logical-OR variant to reduce qubit connectivity.*

The Hamiltonian terms implementing the `found` variable computation may be crafted by pair-wise reduction, introducing more ancillary qubits into the QUBO but with a reduction in the number of highly connected qubits. A comparison of the qubit connection histogram for searching a 100-element array of 8-bit integers, for both the summation and logical-OR variants, is shown in Figure 4.

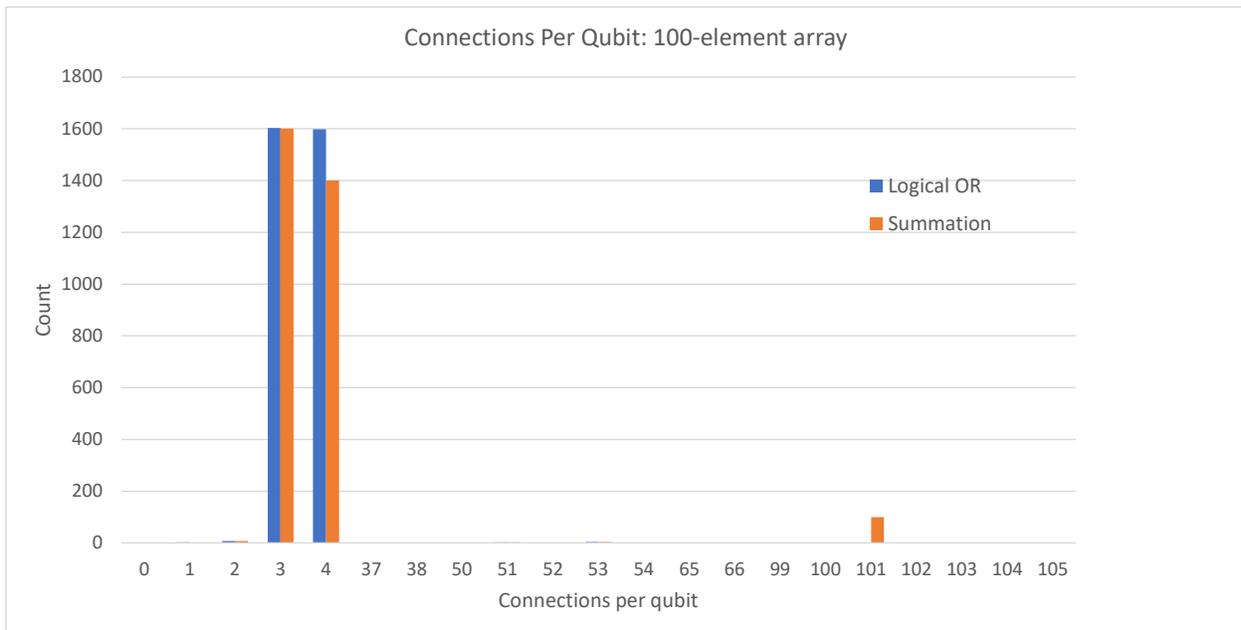

*Figure 4 Histograms of logical qubit connectivity, expressed as connections per qubit, for the summation (orange) and logical-OR (blue) variants of the $H_{search}$ term for a 100-element array. (Truncated for clarity, excluding the 8 counts at 300 connections/qubit common to both options.)*

While more logical qubit variables are required to achieve the same functionality with the logical-OR variant, these additional qubits have 4 connections each instead of the N+1 connections in the summation variant.

**Searches with Multiple Matches**

The purpose of specifying both the index match and value match terms ($I_i$ and $V_i$, respectively) in the treatment above is to avoid constraint violation complications arising when multiple array entries have the sought value. This leaves, however, the question as to which array index is selected if multiple entries satisfy the search criteria. On a well-calibrated machine, all valid solution indices are equally likely in the final system readout.

It is also possible, with only a slight variation in the Hamiltonian, to instead return a count of all matching array entries. The H$_{search}$ constraint term is removed, and instead a counting expression corresponding to

$$count := \sum_{i=1}^{N} V_i$$

where `count` is a quantum integer variable, is added.

**Multiple Criteria Searches**

The search Hamiltonian terms constructed above implemented the comparison of equality between the array element and the sought value, but this is not the only type of search supported. Expanding the comparison term to include more complex expressions is possible, such as range comparisons for simple variable types and composite comparison terms for variable types with multiple fields, by expanding the definition of the $V_i$ term to accommodate the desired comparison logic, e.g.

```
        Vᵢ := (A[i].data_samples > 100) AND (A[i].variance < 0.01 )
```

The H$_{search}$ term is then constructed as in prior sections of this paper.

**A Search Algorithm for Ordered Data Sets**

It is an interesting characteristic of the search algorithm presented that the initial ordering of the data is unimportant to the search performance, e.g., number of logical qubits required. Attempts to find quantum annealing algorithms that leverage *a priori* knowledge of the sorted nature of the data set have thus far proven unfruitful.

The complicating factor is that, for a purely quantum annealing solution, there is a single execution step. While it is possible to replicate sequential operations in a single Hamiltonian, it appears the creation of the superposition state involving comparisons to all array entries is a necessary condition for the search of the array – thereby creating a lower bound on the number of qubits and qubit connections needed for searches contained in a single Hamiltonian.

The result is, we conjecture, there is no improvement possible (in terms of number of qubits required or qubit connectivity) for sorted search versus unsorted search in cases where the data values are determined during the execution of the quantum annealing application.

**Bounding Indices in Ordered Data Sets**

There are many cases where it is useful to find the bounding indices *i* and *i+1* in a sorted data set such that:

```
A[i] <= x < A[i + 1]
```

for some value x.  This can be performed relatively easily on quantum annealing systems by the following algorithm.

The initial step is the creation of comparisons between the value of interest *x* and each value in the data set, storing each comparison result in a 1-qubit variable.

```
C_i := A[i] > x
```

such that $C_i$ is 1 if the data value for A[i] is greater than x and 0 otherwise.

Because we are considering fixed-length arrays in this analysis, the creation of the Hamiltonian terms for these comparisons can be performed explicitly at compile time and does not require the expensive look-up terms required for the unordered data search.

The number of ancillary qubits required for the comparisons is dependent on the data type, for example the comparison of quantum integers of size $K_v$ requires approximately 2 $K_v$ qubits.

Next, a 1-qubit variable is assigned to each 'span' between array elements in the data set, such that the value is 1 if the value x is contained in the i[th] span and 0 otherwise.  This identifies the bounding indices we seek, and is achieved via:

```
span_i := (NOT C_i) AND (C_i+1)            for i ∈ [1..N-1]
```

The Hamiltonian terms for the computation of each `span_i` qubit requires an additional ancillary qubit for order reduction on quantum annealing systems supporting only 2-local terms.  Thus, the algorithm requires approximately $3N + O(K_v)$ qubits to implement.

Limit checking is automatically provided by this approach, with the results presented by the values $C_1$ and $C_N$. If the value x is below all elements in the data set, then $C_1 = 1$, and similarly the upper out-of-bounds condition is reflected by $C_N = 0$.  The `span_i` variables are 0 in these cases since the data is not contained within any span in the array.

**A Sorting Algorithm**

As with the prior sections on data search algorithms, this section on data sorting algorithms considers the case in which the data set to be sorted is populated (in whole or in part) during the run of the quantum annealing application. We assume that data sets whose values are completely known prior to Hamiltonian creation may be sorted, if desired, on digital computing systems prior to the quantum compilation stage.

Because the quantum annealing process is, for simple non-hybrid approaches, a single step, the computation of data elements and the sorting of the data array are not sequential operations. Instead, these 'operations' are achieved simultaneously in the minimization of the system

Hamiltonian, as represented in Figure 5. This has numerous implications driving differences with the digital computing approach, including:

- there is no possibility of an in-place array sort
- the expense of the sort is not expressed as a number of operations

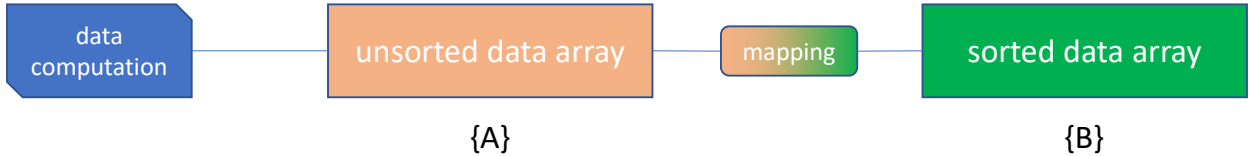

*Figure 5 Notional representation of the simultaneous computation and sorting of data in a quantum annealing application.*

The lack of in-place sort possibilities necessitates the creation of a second array {B} of the same size as the unsorted array {A}. Therefore, upon readout of the final system state, both the sorted and unsorted arrays are present.

The process of sorting is rephrased as a constrained mapping between the unsorted and sorted data arrays, such that:

- The mapping between {A} and {B} is one-to-one.
- The elements of {B} are ordered, according to the specified comparison operation

To implement the mapping, we create a matrix M as an N x N array of 1-qubit variables subject to the one-to-one (into and onto) constraint, which yields the following contribution to the problem Hamiltonian:

$$H_{mapping} = \sum_{j=1}^{N}\left(1 - \sum_{i=1}^{N} M_{ij}\right)^2 + \sum_{i=1}^{N}\left(1 - \sum_{j=1}^{N} M_{ij}\right)^2$$

*Equation 4 The mapping term in the Hamiltonian, ensuring that each index in {A} is mapped to exactly one index in {B}, and each index in {B} maps to exactly one index in {A}.*

The mapping must then be applied, so that the elements of {A} are replicated in {B}. This requires $N^2$ 'operations' of the form:

```
if ( M_ij == 1) then
    B[j] := A[i]
end if
```

Expressed as a portion $H_{assign}$ of the system Hamiltonian, and assuming bit-wise copying of the data values, this becomes:

$$H_{assign} = \sum_{i=1}^{N}\sum_{j=1}^{N} M_{ij} \left[\sum_{l=1}^{K_v}(A[i]_l + B[j]_l - 2\,A[i]_l B[j]_l)\right]$$

*Equation 5 The Hamiltonian phrase to ensure elements of array {B} correspond to the mapped entries from array {A}. Terms for bit-wise equivalence contribute to the total energy if and only if the mapping $M_{ij}$ was selected.*

The assignment portion of the Hamiltonian contains 3-local terms, necessitating $K_v N^2$ ancillary qubits for order-reduction to run on systems whose hardware only supports 2-local terms in the Hamiltonian.

To implement the ordering of the array {B}, we have N – 1 constraints of the form:

$$B[i] \leq B[i+1] \quad for\ i\ \in [1..N-1]$$

*Equation 6 Sorted array element constraint.*

The exact form of these constraints will depend on the data type, and generally scales as $O(K_v)$ in terms of the number of qubit variables required for enforcement. The combination of the mapping, assignment, and comparison constraints expresses a consistent system whereby the data elements of unsorted array {A} are present in sorted array {B}.

As with the case of the bounding index search, these are not random-access array lookups. While the values of the array are not known at compile time, the order of the indices in the sorted array {B} are known, thereby permitting the explicit construction of each term in Equation 6 at compile time. This avoids the expensive lookup seen in searching unordered data sets.

Hamiltonians were constructed to sort arrays for which the data element types are 8-bit quantum integers. Examination of the connectivity of the qubit graph representing the mapping and assignment terms reveals a growth slightly above linear as a function of array size (Figure 6) for the most-connected qubit in the problem. As this quickly exceeds the hardware capabilities of current systems, the connectedness (and thus the problematic embedding) will likely prove the limiting factor for including in-application quantum annealing sorting for the near term.

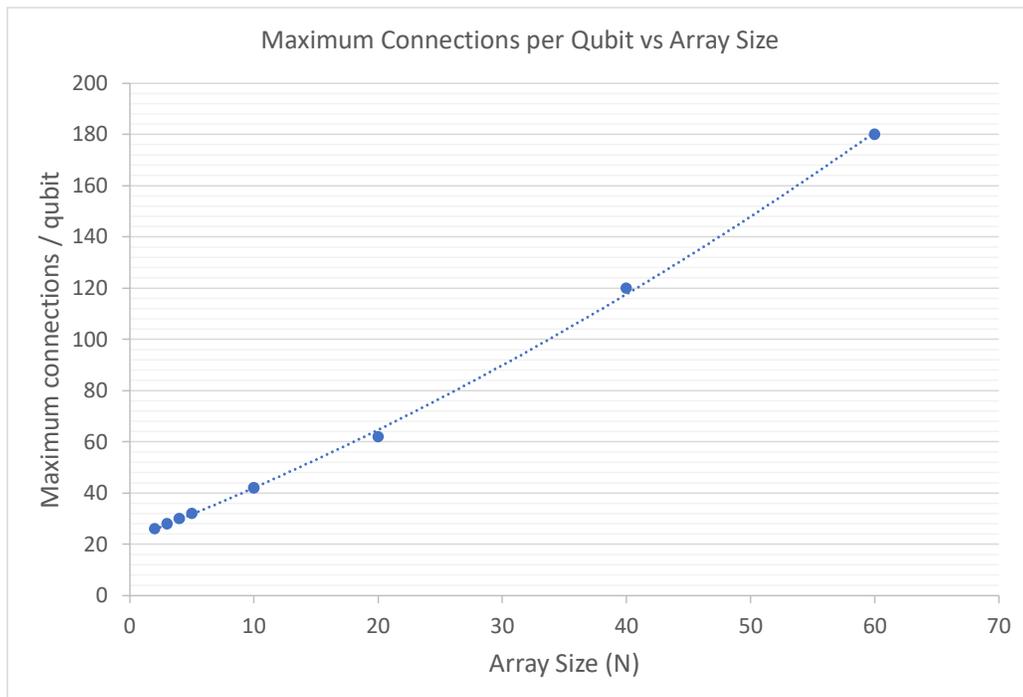

*Figure 6 Maximum connections per qubit vs array size, for sorting arrays of 8-bit quantum variables.*

The number of qubit variables in the Hamiltonian required to implement the mapping and assignment terms for the arrays of 8-bit quantum variables ($K_v = 8$), shown in Figure 7, increases as expected and is dominated by the $K_v N^2$ contribution from the order-reduction qubits in $H_{assign}$.

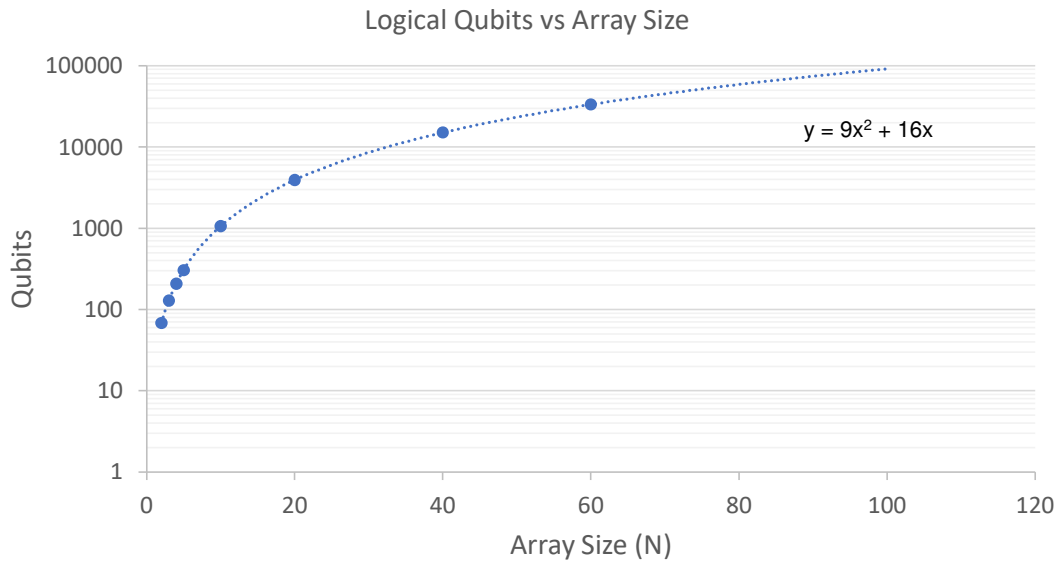

Figure 7 The number of logical qubits (variables and ancillary) required in $H_{mapping}$ and $H_{assign}$ to sort an array of 8-bit quantum variables, as a function of array size. The fit curve shows the $K_v N^2$ from the order reduction qubits and the $N^2$ qubits from the M mapping matrix driving the leading term of $9 N^2$.

**Sorting Ordered Data Sets**

Some digital computing sorting algorithms (e.g., quicksort) have notoriously poor performance when the input data are nearly or completely sorted. The sorting algorithm presented here is agnostic with respect to the ordering of the input data. As with the search case, it is the size and variable type of the data set, rather than the ordering, that determine the number and connections of the qubits needed to implement the sorting algorithm for quantum annealing systems.

**Stability in the Sorting Algorithm**

A sorting algorithm is considered 'stable' if the relative order of equivalently ranked data elements is unchanged when sorting. The sorting algorithm presented here is not a stable sort and does not guarantee preservation of order. In these cases of equivalent sort order, there are degenerate ground states in the Hamiltonian – each of which is a valid solution – and on well-calibrated hardware will be equally likely in the final readout.

**Preprocessing Effort**

As with the search option, there is no special preprocessing necessary for the array data; all preprocessing costs are in the construction of the Hamiltonian terms for the sort. Generation of the $H_{mapping}$ terms requires $O(N^3)$ steps, and the generation of the $H_{assign}$ terms requires $O(K_v N^2)$ steps, thereby giving a total digital computer preprocessing effort of $O(N^3)$ steps to build the Hamiltonian terms implementing the sort algorithm.

**Conclusions**

The non-sequential nature of quantum annealing and adiabatic quantum computing systems may not appear conducive to operations such as searching and sorting, which are inherently sequential in their digital computing implementations. However, in this paper we have shown that the construction of Hamiltonian terms implementing such functionality is a straight-forward matter with well-characterized scaling as a function of array size.

Searching arrays whose data values are created during the execution of the quantum application, accessing array entries where the array index is determined during the execution of the quantum application, searching arrays for a value created during the execution of the quantum application, and finding the bounding indices of data arrays have been presented in algorithms that scale linearly, as a function of array size, in both the number of logical qubits required and the connectivity per qubit. At this time, the qubit connectivity is likely to be the limiting factor in the utilization of array searches in quantum annealing applications given the limitations of current hardware qubit graphs.

The algorithms provided for array searches handle the case of multiple data matches, albeit by means of a Hamiltonian with degenerate, equally likely ground states.

We find that, in the case of run-time determined array values, the necessity of comparing the target search value against every array element drives a lower bound in the number of comparison operations that is independent of the ordering of the array values. We conjecture that searching sorted arrays of quantum variables, in the case of quantum annealing systems, can be no more efficient (in terms of system resources) than searching unsorted arrays of quantum variables.

We have also presented an algorithm for the sorting of arrays of quantum variables, whose resource demands grow quadratically as a function of array size. This is in contrast with known sorting algorithms for digital computers that can operate in $O(N \log N)$ steps and gate array quantum computers algorithms in $\Omega(N \log N)$ steps [2]. While it was not the goal of this analysis to construct a quantum algorithm of performance superior to that of digital computing architectures, the disparity raises the possibility that the quantum algorithm presented herein may not be the most efficient possible. Nonetheless, given the stated goal of creating an algorithm for use within the context of larger quantum computing applications, this first delivery achieved the desired capability.

**Appendix A: The Equality Comparison of Composite Quantum Variables**

An initial version of array element comparison suitable for integer quantities has been previously published [4], so presented below is the more general version suitable for composite quantum variables of any type for which equivalency can be determined by bitwise comparison – including fixed-length floating point numbers, strings, etc.

Two such variables are considered equal if and only if they are bitwise equal, i.e.

$$equality := \bigwedge_{all\ bits\ j} X_j == Y_j$$

Unlike bitwise assignment, for which the Hamiltonian term is easily constructed by the addition of $(X_j - Y_j)^2$ terms for each bit, we seek to determine the equivalency and thus require assignment to a single qubit of the results of the bit comparison. This is slightly more complex in that it requires the introduction of ancillary qubits to store intermediate results and for order reduction, but the technique is straight-forward.

For each bit, we perform an initial determination:

$C_j := (X_j == Y_j)$

which mathematically is

$C_j = 1 - (X_j - Y_j)^2$

and corresponds to a Hamiltonian term of the form:

$2 X_j Y_j - X_j - Y_j - C_j + 2 X_j C_j + 2 Y_j C_j - 4 X_j Y_j C_j$

which may, with the addition of an ancillary qubit, be reduced to 2-local form by standard techniques.

After the construction of the bitwise match variables, $C_j$, the equality check for the composite variable becomes:

$$equality := \bigwedge_{all\ bits\ j} X_j == Y_j \ = \bigwedge_{all\ bits\ j} C_j$$

This expression may be simplified by pairwise reduction of the variables (via ancillary qubits) to the form:

$equality := A\ AND\ B$

the Hamiltonian for which is well-documented in the literature, via the gadget for $z = xy$.

In this manner, a generalized Hamiltonian expression for the equality determination of two composite quantum variables whose values are not known at compile time may be constructed. This is the technique used for this paper, and the needed ancillary qubits are included in the problem qubit sizes quoted in the scalability analyses.